\documentclass{moriond}

\usepackage{amsmath} 
\def\be{\begin{equation}}
\def\ee{\end{equation}}
\def\bea{\begin{eqnarray}}
\def\eea{\end{eqnarray}}

\newcommand*{\ifb}{\mbox{fb$^{-1}$}}

\newcommand{\Photo}{}

\begin{document}
\vspace*{4cm}
\title{Higgs couplings in ATLAS at Run2}

\author{ H.M. Gray on behalf of the ATLAS Collaboration }

\address{Department of Physics, University of California, Berkeley, 425 Le Conte Hall, Berkeley, CA 94720}

\maketitle\abstracts{
Since the discovery of the Higgs boson in summer 2012, the understanding
of its properties has been a high priority of the ATLAS physics program.
Measurements of Higgs boson properties sensitive to its production processes,
decay modes, kinematics, mass, and spin/CP properties based on $pp$ collision
data recorded at 13 TeV are presented. The analyses of several production
processes and decay channels will be described, including recent highlights
as the direct observation of the couplings to top and beauty quarks,
and an updated combination of all measurements.
}

\section{Introduction}
It is now ten years after the first beam from the Large Hadron Collider (LHC) was observed in the ATLAS experiment and seven years from the discovery of the Higgs boson by the ATLAS and CMS experiments in 2012~\cite{HIGG-2012-27,CMS-HIG-12-028}. However, the current dataset is only 5\% of the total dataset expected from the upgraded LHC, the high-luminosity LHC (HL-LHC), which is expected to start delivering data in seven years. 

Over the seven years since it was discovered, many properties of the Higgs boson have been established. The mass has been measured using a combination of the Run-1 and Run-2 data to be $124.97 \pm 0.24$~GeV, reaching a precision of 0.2\%~\cite{HIGG-2016-33}. The width has been constrained to be less than 14.4~MeV (15.2~MeV expected) via a measurement of off-shell $ZZ$ production~\cite{HIGG-2017-06}. Measurements of the spin and parity have been found to be consisted with the value expected from the Standard Model: $J^{PC} = 0^{++}$~\cite{HIGG-2013-17}. This article will review the current status of the measurements by the ATLAS experiment of the couplings of the Higgs boson to the particles of the Standard Model. The measurements discussed either use the dataset collected by ATLAS during 2015 and 2016 with an integrated luminosity of $36~\ifb$ or also include the data collected during 2017 with a total integrated luminosity of $80~\ifb$.

\section{Probing Higgs Couplings at the LHC}
There are four main production modes for Higgs boson production at the LHC. The dominant production channel is gluon-gluon fusion, which typically proceeds via a loop containing top quarks. The vector boson fusion (VBF) production channel is characterised by two forward jets and little central hadronic activity. Higgs boson production in association with a vector boson (VH) is identified by tagging the $W$ or $Z$ boson. Finally, production of the Higgs boson in association with top quarks is identified by tagging the two top quarks. The cross-sections and the number of Higgs bosons produced during Run-2 of the LHC are summarised in Table~\ref{tab:prod}. There are five key decay channels at the LHC ordered by decreasing branching ratio: $b$-quarks (58\%), $W$-bosons (21\%), $\tau$-leptons (6.3\%), $Z$-bosons (2.6\%), and photons (0.23\%)~\cite{Heinemeyer:1559921}. Conveniently for experimentalists, the mass of the Higgs boson happens have the value for which the largest number of decay modes can be probed experimentally.

\begin{table}[t]
\caption[]{Main production modes for the Higgs boson at the LHC. The cross-section at a centre-of-mass energy of 13~TeV is shown in the second column. The number of Higgs bosons produced in each channel during Run-2 at the LHC is shown in the third column. From Heinemeyer \textit{et al.}~\cite{Heinemeyer:1559921} and references therein.}
\label{tab:prod}
\vspace{0.4cm}
\begin{center}
\begin{tabular}{|l|c|c|}
\hline\hline
Name & Cross-section [pb] & \#Higgs bosons \\
\hline
Gluon-gluon fusion (ggF) & 49 & 6.9M \\
Vector boson fusion (VBF) & 3.8 & 520k \\
Vector-boson associated production (VH) & 2.3 & 320k \\
Top-quark associated production (ttH) & 0.5 & 70k \\
\hline\hline
\end{tabular}
\end{center}
\end{table}

\section{Coupling of the Higgs to Bosons}
Measurements of the discovery channels, $H \rightarrow \gamma\gamma$~\cite{ATLAS-CONF-2018-028} and $H \rightarrow ZZ \rightarrow 4l$~\cite{ATLAS-CONF-2018-018}, have been performed using 80~\ifb of data. As shown in Figure~\ref{fig:HZZHgg}, the current results probe the cross-sections of the individual production modes with precision reaching 15\%. At present, both the experimental and theoretical systematic uncertainties play a key role.

In the $H \rightarrow WW$ channel, cross-section measurements have been made for both ggF and VBF production~\cite{HIGG-2016-07}. For ggF production, the observed cross-section is $11.4\ ^{+1.2}_{-1.1} \mathrm{\ (stat.)}\ ^{+1.2}_{-1.1} \mathrm{\ (theo\ syst.)} \\ ^{+1.4}_{-1.3} \mathrm{\ (exp\ syst.)}~\mathrm{pb}$. This is in good agreement with the SM theoretical prediction of $10.4 \pm 0.6$~pb. Similarly, for VBF production, the observed cross-section is $0.5\ ^{+0.24}_{-0.22} \mathrm{\ (stat.)}\ \pm\ 0.10 \mathrm{\ (theo\ syst.)}\ ^{+0.12}_{-0.13} \\ \mathrm{\ (exp\ syst.)}~\mathrm{pb}$. This is in good agreement with the SM theoretical prediction of $0.81 \pm 0.02$~pb.

\begin{figure}[hbtp!]
\begin{minipage}{0.49\linewidth}
\centerline{\includegraphics[width=\linewidth,]{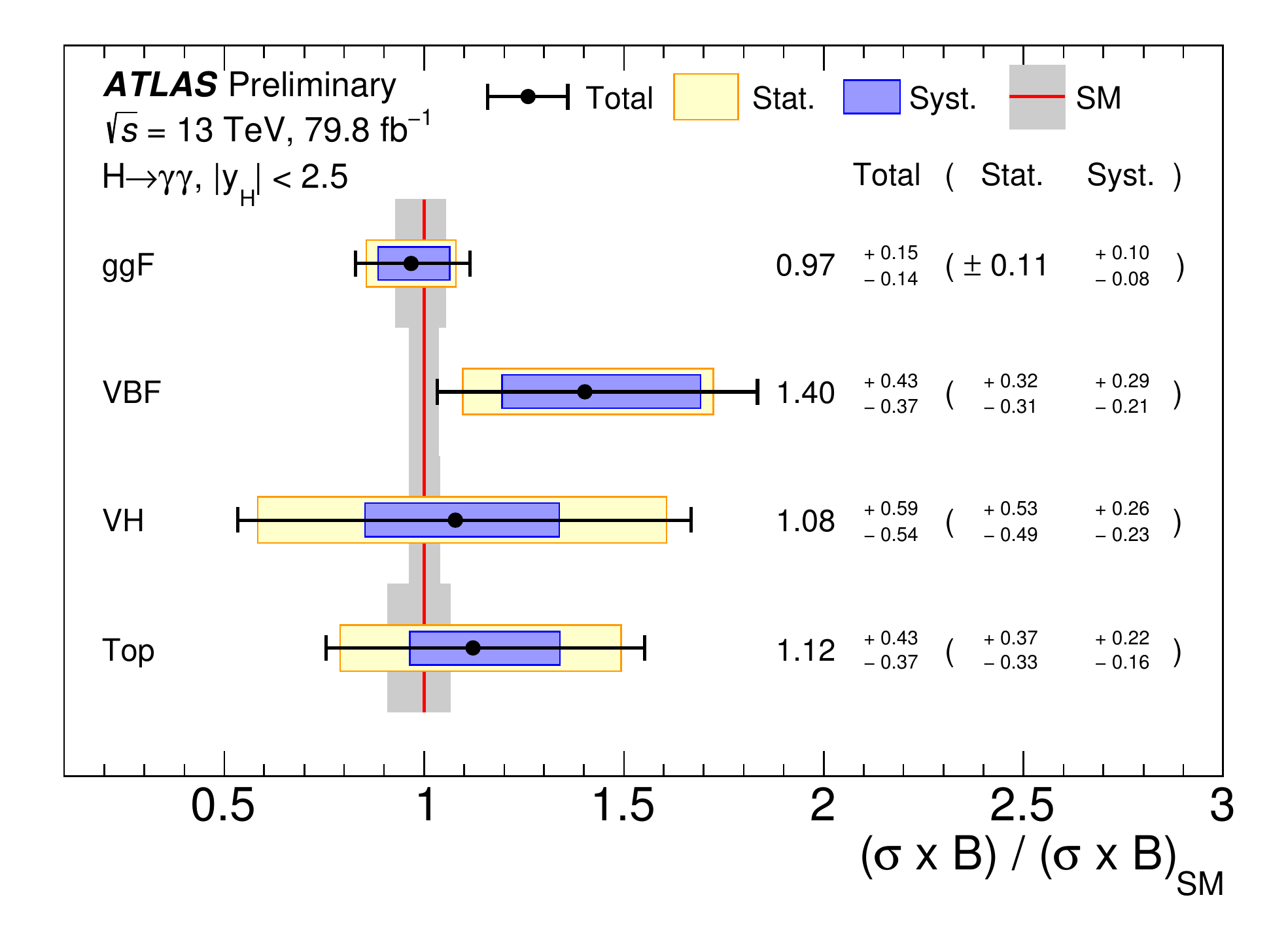}}
\end{minipage}
\hfill
\begin{minipage}{0.43\linewidth}
\centerline{\includegraphics[width=\linewidth]{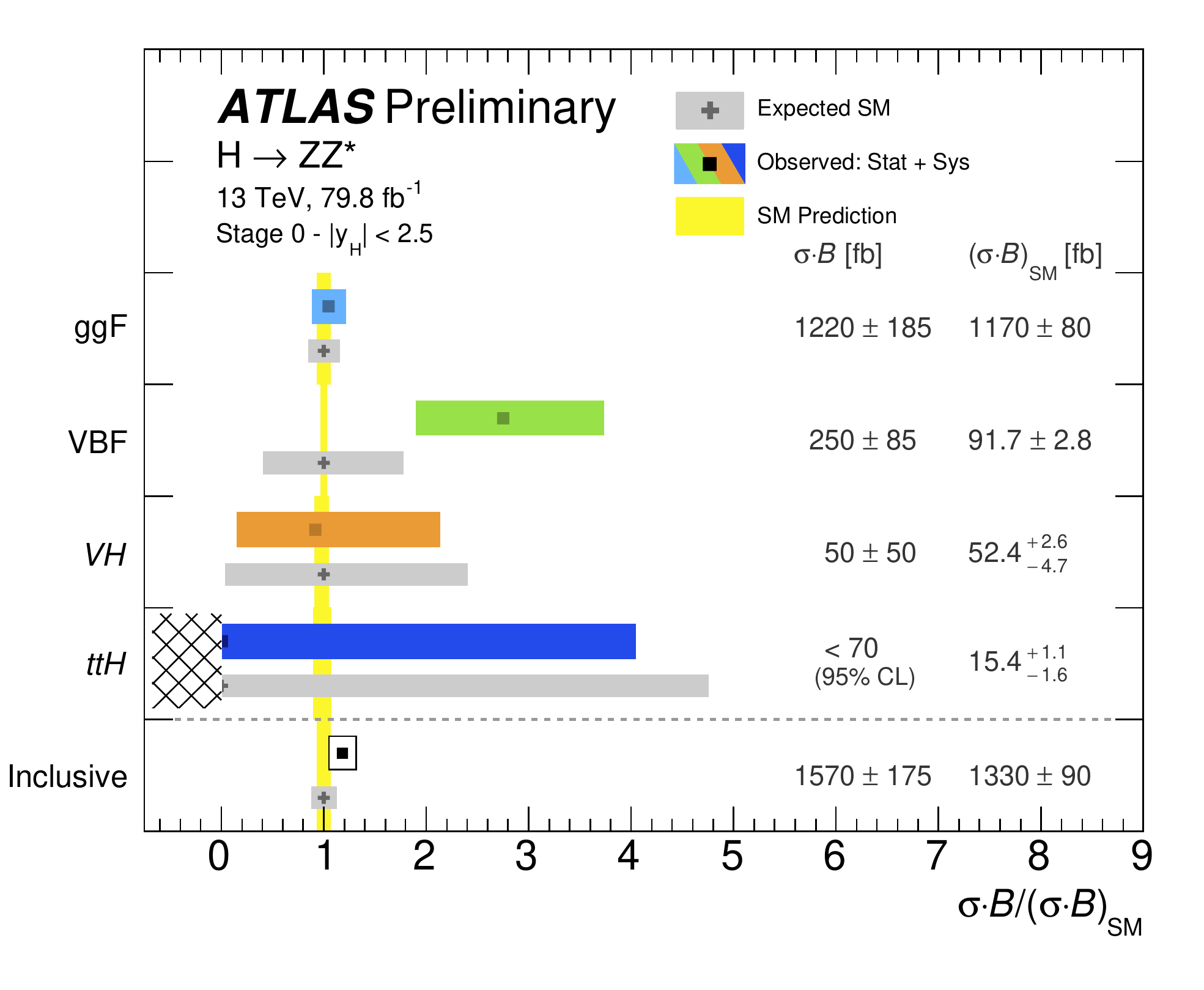}}
\end{minipage}
\hfill
\caption[]{The measured cross-sections for different Higgs boson production modes as measured by ATLAS in the $H \rightarrow \gamma\gamma$~\cite{ATLAS-CONF-2018-028} (left) and $H \rightarrow ZZ* \rightarrow 41$~\cite{ATLAS-CONF-2018-018} (right) channels. }
\label{fig:HZZHgg}
\end{figure}

\section{Observation of the coupling of the Higgs to fermions}
The first observation of the coupling of the Higgs boson to fermions was made in the combination of the ATLAS and CMS Run-1 results~\cite{HIGG-2014-14} reaching an observed (expected) significance of $5.5 (5.0) \sigma$. ATLAS has achieved an observed (expected) significance of $6.4 (5.4)\sigma$ of the decay of the Higgs boson to $\tau$'s by combining the Run-1 results with a Run-2 analysis using $36~\ifb$~\cite{HIGG-2017-07}. In addition, the production cross-sections for ggF and VBF have been measured for the $H \rightarrow \tau\tau$ decay channel. For ggF production, the observed cross-section is $3.1\ \pm 0.1 \mathrm{\ (stat.)}\ ^{+1.6}_{-1.3} \mathrm{\ (syst.)}~\mathrm{pb}$, which in in good agreement with the SM theoretical prediction of $3.05 \pm 0.13$~pb. Similarly, for VBF production, the observed cross-section is $0.28\ \pm 0.09 \mathrm{\ (stat.)}\ ^{+0.11}_{-0.9} \\ \mathrm{\ (exp\ syst.)}~\mathrm{pb}$. This is in good agreement with the SM theoretical prediction of $0.237 \pm 0.006$~pb. This shows that the $H \rightarrow \tau\tau$ channel is a particularly sensitive channel for probing VBF production.

A physics highlight at the LHC during 2018 was the observation of the coupling of the Higgs boson to $b$-quarks~\cite{HIGG-2018-04}. Despite the large branching ratio of 58\% it is a difficult channel due to large backgrounds. The most sensitive production mode is VH production, but ATLAS has also performed searches targetting ggF, VBF and ttH production. Through the combination of all these production modes and using both Run-1 and Run-2 data, ATLAS observed the decay of the Higgs boson to $b$-quarks with an observed (expected) significance of $5.4 (5.5)\sigma$. The VH analysis strategy was cross-checked by the observation of $VZ(bb)$ production. Figure~\ref{fig:fermions} (left) shows the $m_{bb}$ distribution obtained in the VH analysis of the Run-2 data with contribution from $H \rightarrow bb$ shown in red and $Z \rightarrow bb$ in grey. The $VH(bb)$ channel is also the most sensitive channel for VH production.

\begin{figure}
\begin{minipage}{0.49\linewidth}
\centerline{\includegraphics[width=\linewidth]{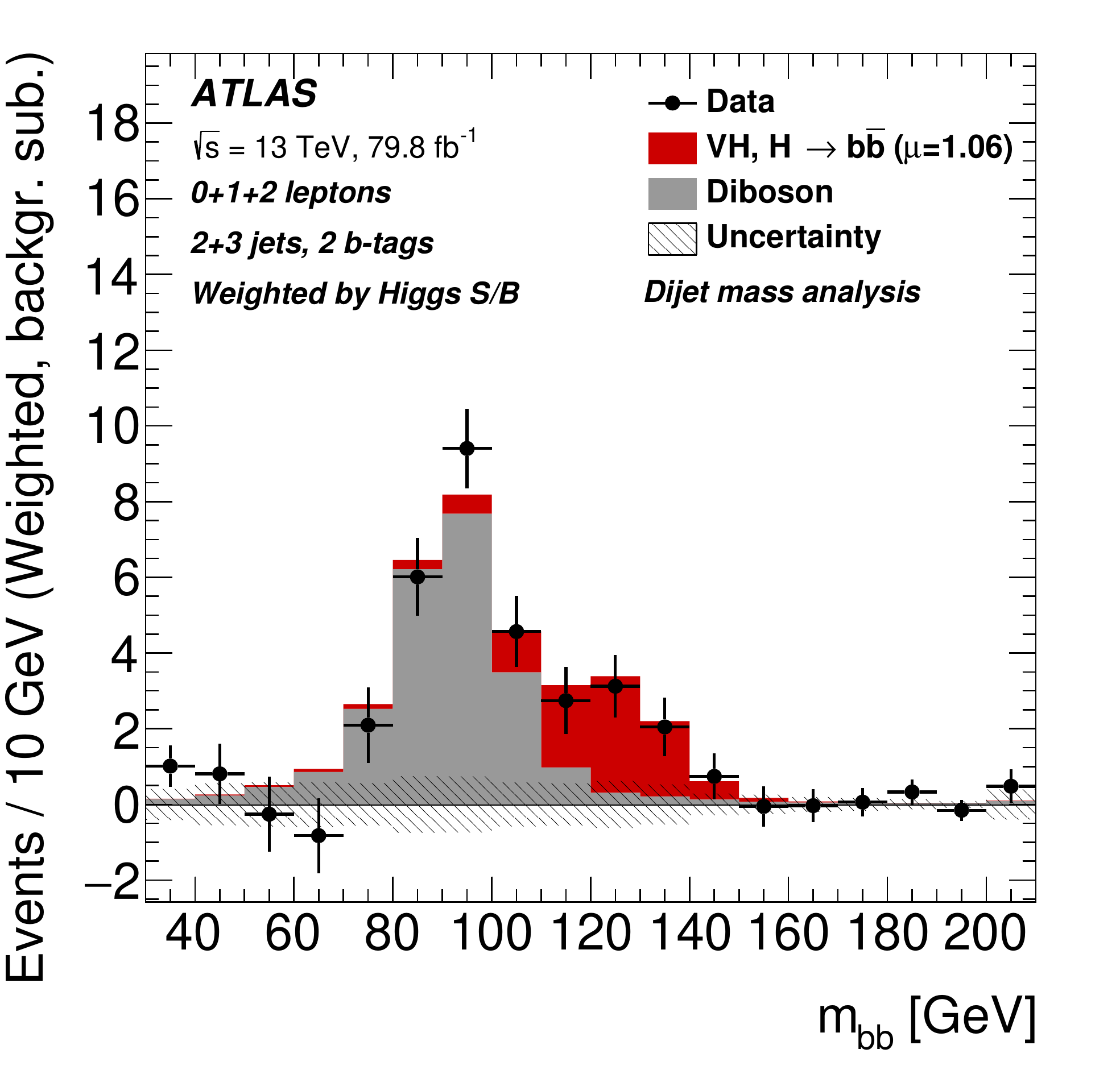}}
\end{minipage}
\hfill
\begin{minipage}{0.48\linewidth}
\centerline{\includegraphics[width=\linewidth]{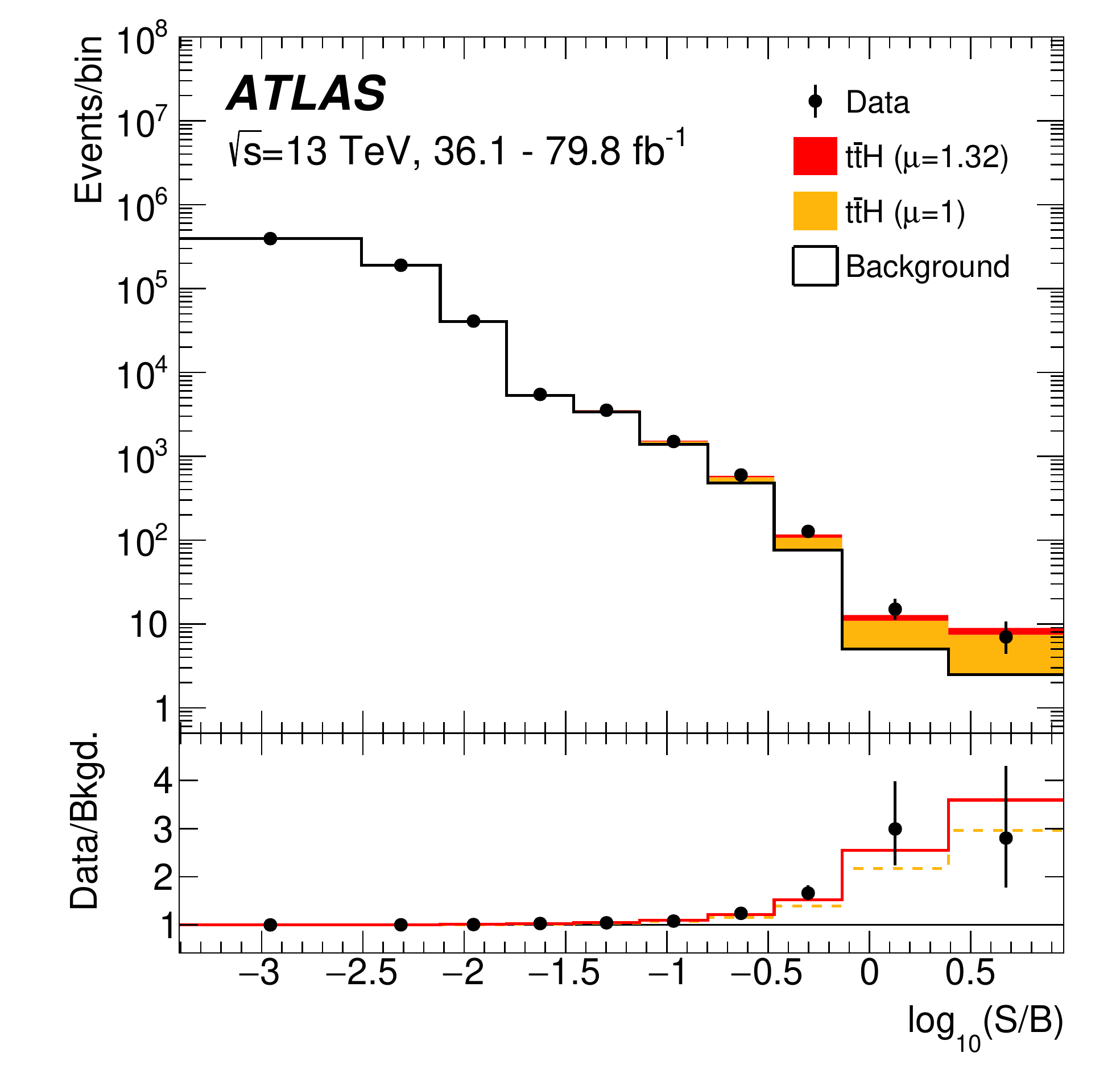}}
\end{minipage}
\hfill
\caption[]{The dijet invariant mass distribution in data from the VH(bb) analysis after all backgrounds except for the WZ and ZZ processes have been subtracted~\cite{HIGG-2018-04} (left). The observed event yields in all categories from the ttH analysis ordered by $log_{10}(S/B)$~\cite{HIGG-2018-13} (right).}
\label{fig:fermions}
\end{figure}

Another highlight during 2018 was the observation of the coupling of the Higgs boson to top quarks~\cite{HIGG-2018-13}. ttH production is particularly interesting because it provides a direct probe of the coupling of the Higgs boson to top quarks, which means that in combination with ggF production, it probes potential new physics contributions within the ggF loop. Through the combination of all major decay modes, ATLAS observed ttH production with an observed (expected) significance of $6.3 (5.1)\sigma$. Figure~\ref{fig:fermions} (right) shows the analysis bins used in the combination ordered by increasing signal over background. The expected contribution from SM ttH production is shown in orange and the observed contribution in red.

\section{Rare and Invisible Higgs Decays}
The growing dataset at the LHC has allowed ATLAS to set limits at the 95\% CL on channels targeting decays of the Higgs boson in addition to the five key decay channels. One example is the rare decay of the Higgs boson to a $Z$-boson and a photon, in which ATLAS has set an observed (expected) limit of 6.4 (4.4) times the SM prediction~\cite{HIGG-2016-14}. A key target is probing the coupling of the Higgs boson to fermions outside the third generation. Here, the most powerful channel is $H\rightarrow \mu\mu$ in which ATLAS has set an observed (expected) limit of 2.1 (2.0) times the SM prediction~\cite{ATLAS-CONF-2018-026}. Two analyses have probed the coupling of the Higgs boson to charm quarks: ZH production with the $Z$ boson decaying to leptons and the Higgs boson to charm quarks with a limit 110 times the SM prediction~\cite{HIGG-2017-01} and $H \rightarrow J/\Psi \gamma$ with a limit 120 times the SM prediction~\cite{HIGG-2016-23}. Limits have also been set on $H \phi\gamma$ at 200 times the SM prediction and on $H \rightarrow \rho\gamma$ at 50 times the SM prediction~\cite{HIGG-2016-13}.

Indirect constraints have been set on the decay of the Higgs boson to invisible particles from fits to the combined Higgs coupling results. Direct constraints have been obtained from three ATLAS searches: production in association with a $W$ or $Z$ boson decaying to hadrons; production in association with a $Z$ decaying to leptons and VBF production. In the combination of the three channels, ATLAS set a limit on the branching ratio of the Higgs boson to invisible particles of 26\% at the 95\% CL~\cite{ATLAS-CONF-2018-054}. This is slightly weaker than expected from the SM of $17^{+0.07}_{-0.05} \%$.

\begin{figure}
\begin{minipage}{0.45\linewidth}
\centerline{\includegraphics[width=\linewidth]{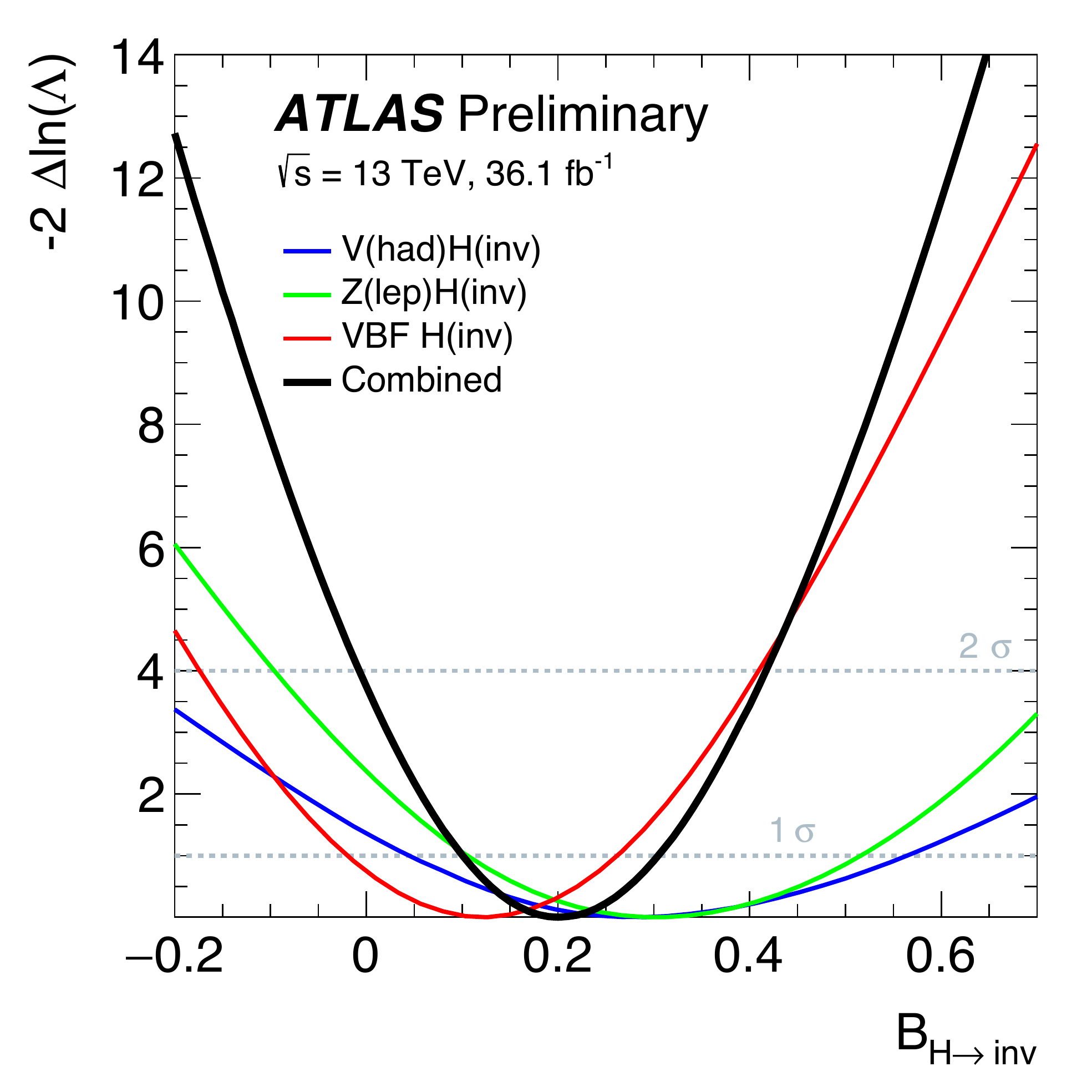}}
\end{minipage}
\hfill
\begin{minipage}{0.48\linewidth}
\centerline{\includegraphics[width=\linewidth]{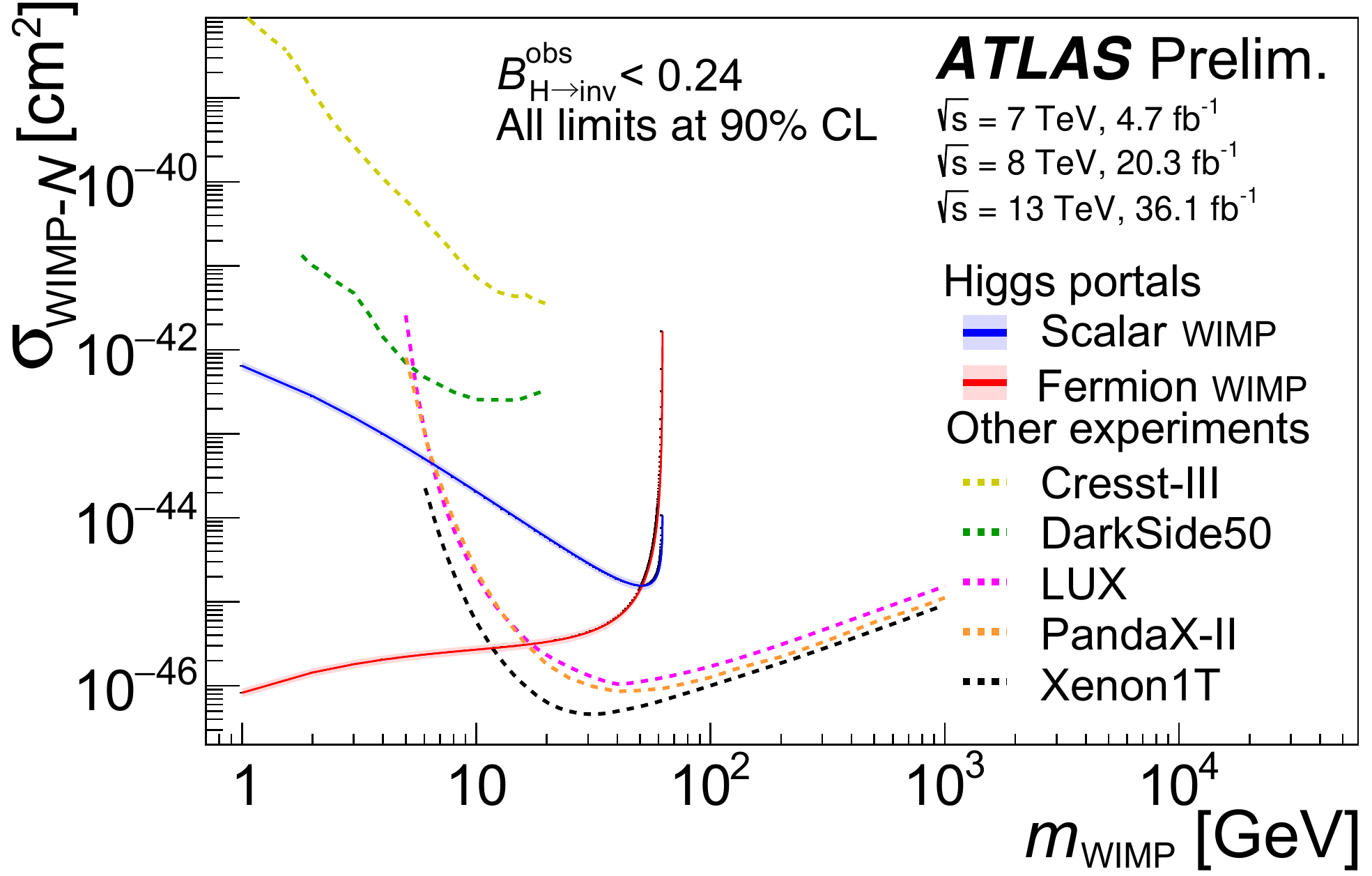}}
\end{minipage}
\hfill
\caption[]{
The observed negative logarithmic profile likelihood ratios as a function of the branching ratio of $H \rightarrow \text{inv}$ of the VH, ZH, and VBF topologies in Run 2 (left)~\cite{ATLAS-CONF-2018-054}. Comparison of upper limits at 90\% CL from direct detection experiment on the spin-independent WIMP-nucleon scattering cross-sections to the exclusion limit from ATLAS~\cite{ATLAS-CONF-2018-054}}
\label{fig:higgsinv}
\end{figure}

\section{Higgs Combination}
ATLAS has performed a new combination~\cite{ATLAS-CONF-2019-005} to extract the properties of the Higgs boson from these results. When the fit is performed to extract a single signal strength parameter, the following value is obtained:
\begin{equation}
\mu = 1.11^{+0.09}_{-0.08} = 1.11 \pm 0.05 \mathrm{(stat.)} ^{+0.05}_{-0.04} \mathrm{(exp.)} ^{+0.05}_{-0.04} \mathrm{(sig. th.)} ^{+0.03}_{-0.03} \mathrm{(bkg. th.)}
\end{equation}

In this combination, significances in excess of $5\sigma$ are obtained for the ggF (not quoted), VBF ($6.5\sigma$), VH ($5.3\sigma$) and ttH ($5.8\sigma$) production modes when assuming SM branching ratios. Low correlations between the different production modes are observed in the fit. Figure~\ref{fig:fit} (left) shows the cross-sections normalized to the SM values for each production mode. The results are consistent with the predictions from the SM.

The combination is also performed to simultaneously extract the product of each major production and decay mode. The production and decay modes with low sensitivity are fixed to the SM values. Again, no significant deviations from the SM predictions are observed. The results are interpreted within the $\kappa$-framework to extract the couplings of the Higgs boson to different SM particles. In this fit, it is assumed that there are no contributions from physics beyond the SM to the total width. Figure~\ref{fig:fit} (right) shows the extracted couplings as a function of particle mass. Good agreement is observed for the couplings of all particles with respect to the predictions from the SM.

\begin{figure}[hbtp!]
\begin{minipage}{0.49\linewidth}
\centerline{\includegraphics[width=\linewidth,]{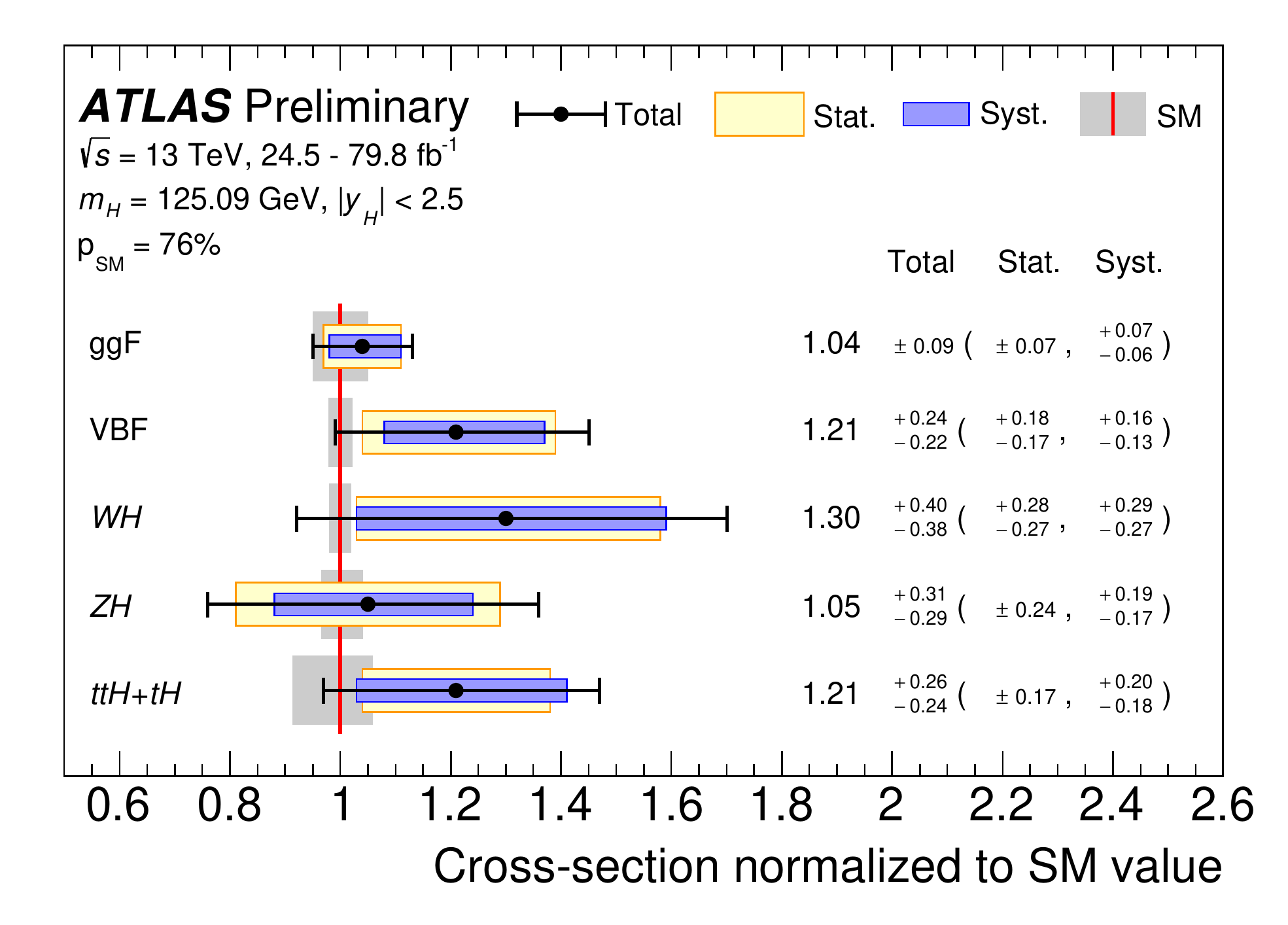}}
\end{minipage}
\hfill
\begin{minipage}{0.48\linewidth}
\centerline{\includegraphics[width=\linewidth]{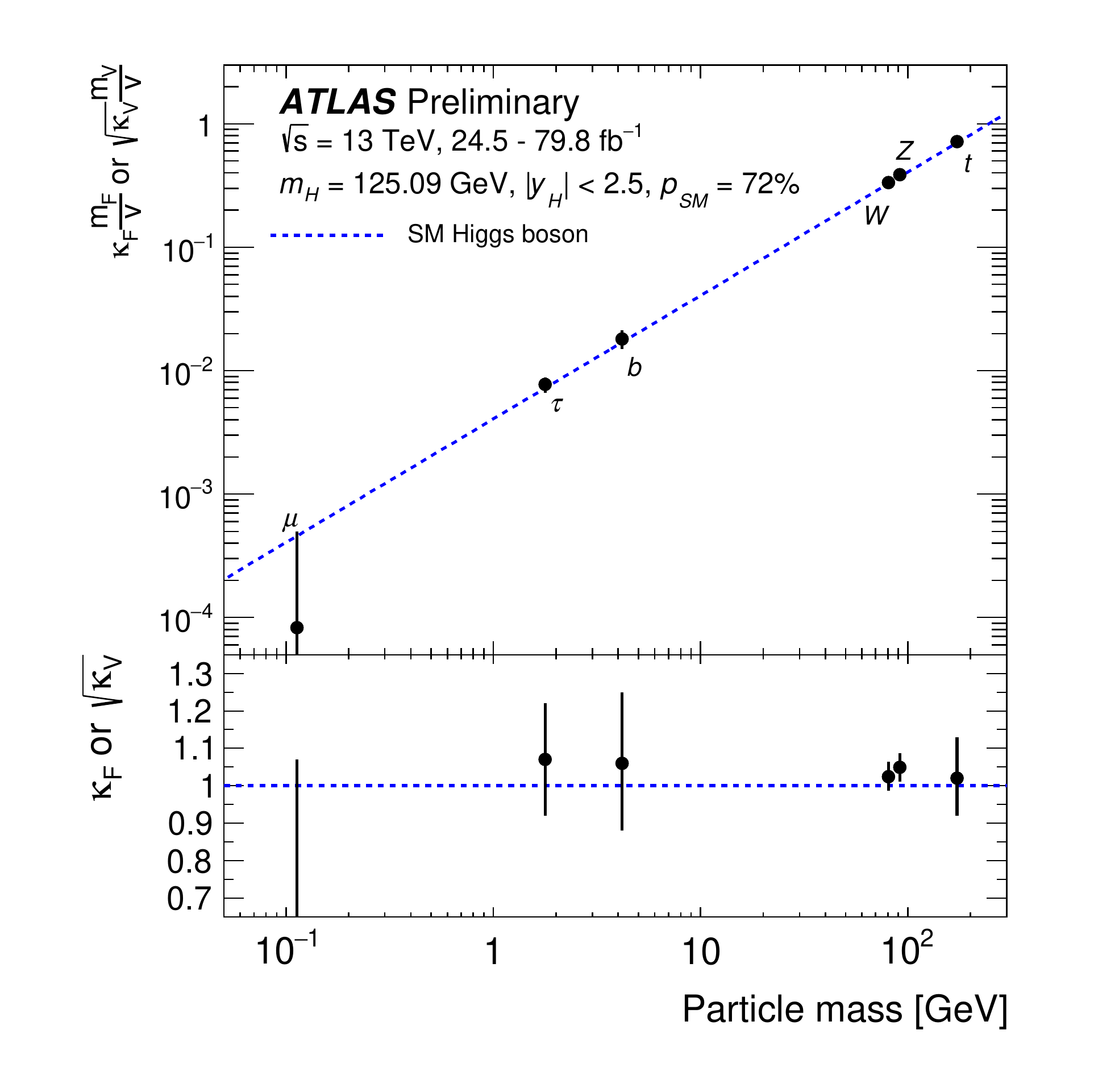}}
\end{minipage}
\hfill
\caption[]{The measured cross-sections for Higgs boson production from the ATLAS combination~\cite{ATLAS-CONF-2019-005}. The measured coupling strength as a function of particle mass~\cite{ATLAS-CONF-2019-005}.}
\label{fig:fit}
\end{figure}

To probe potential BSM contributions, fits have been performed using three different assumptions on the total width of the Higgs boson. The first assumes that the branching ratio to BSM particles is zero. The second constrains the coupling for vector bosons, $\kappa_V$, to be less than one. The third assumes that the on-shell and off-shell values for $\kappa$ are identical. In all cases, the measured scale factors are found to be compatible with the predictions from the SM. In the second scenario, an observed (expected) limit of the branching ratio of the Higgs boson to invisible particles of 30\% (18\%) and an observed (expected) limit of the branching ratio of the Higgs boson to undetected particles of 22\% (38\%). In the third scenario and largely due to the off-shell $H \rightarrow ZZ \rightarrow 4l$ measurement, an observed (expected) limit on the branching ratio of the Higgs boson to BSM particles of 47\% (57\%) is set~\cite{ATLAS-CONF-2019-005}. 

\section{Probing the Higgs Self-Coupling}
The measurement of the Higgs self-coupling is an important physics target for the (HL)-LHC. In addition to direct searches looking for diHiggs production, constraints can be set from single Higgs production cross-sections. Following Degrassi \textit{et al.}~\cite{Degrassi:2016wml} and Maltoni \textit{et al.}~\cite{Maltoni:2017ims}, ATLAS starts from the Higgs potential in the following form:
\begin{equation}
 V(H) = \frac{1}{2}m_H^2 + \lambda_3\nu H^3 + \frac{1}{4}\lambda_4H^3 + O(H^5)
\end{equation}
and consider the following modifications: $\lambda_3 = \kappa_{\lambda} \lambda_3^{SM}$, where $\kappa_{\lambda}$ scales the strength of the Higgs self-coupling with respect to the prediction from the SM.

This results in a modified Higgs boson production cross-section:
\begin{equation}
 \mu_i(\kappa_{\lambda}, \kappa_i) = \frac{\sigma^{BSM}}{\sigma^{SM}} = Z_H^{BSM}(\kappa_{\lambda})\left[ \kappa_i^2 + \frac{(\kappa_{\lambda} - 1) C_1^i}{K_{EW}^i} \right]
\end{equation}
where $Z_H^{BSM}(\kappa_{\lambda}) = \frac{1}{1 - (\kappa_{\lambda}^2 - 1) \delta Z_H}$ with $\delta Z_h = -1.536 \times 10^{-3}$ and modified branching fractions:
\begin{equation}
 \mu_f(\kappa_{\lambda}, \kappa_f) = \frac{BR_f^{BSM}}{BR_f^{SM}} = \frac{\kappa_f^2 + (\kappa_{\lambda} - 1)C_1^f}{\sum_j BR_j^{SM} \left[ \kappa_j^2 + (\kappa_{\lambda} - 1)C_1^j \right]}
\end{equation}

This parametrization was applied to the production cross-sections and branching fractions and the full Higgs combination fit including the simplified template cross-section results was performed~\cite{ATL-PHYS-PUB-2019-009}. The following value for $\kappa_{\lambda}$ was obtained:
\begin{equation}
\kappa_{\lambda} = 4.0^{+3.7}_{-3.6} \mathrm{(stat.)} ^{+1.3}_{-1.5} \mathrm{(exp.)} ^{+1.3}_{-0.9} \mathrm{(sig. th.)} ^{+0.8}_{-0.9} \mathrm{(bkg. th.)}
\end{equation}
Alternately, when a limit is set at the 95\% confidence level, it is found to lie in the following interval: $-3.2 < \kappa_{\lambda} < 11.9$.

\section{Conclusion}
The status of the current results from the ATLAS experiment for the measurement of the coupling of the Higgs boson using the Run-2 dataset has been presented. Highlights since Moriond 2018 include the observation of the coupling of the Higgs to bottom quarks, observation of the coupling of the Higgs to top quarks and a new combination with the observation of all main Higgs production modes at the LHC. So far, the results seem to agree to a remarkable degree to the predictions from the Standard Model. However, there is more data and many more results still to come.

\section*{References}
\bibliography{bibliography} 

\end{document}



